\documentclass[journal=nalefd,manuscript=article]{achemso}
\setkeys{acs}{
email = false
}
\usepackage[utf8]{inputenc}
\usepackage{mathtools}
\usepackage{graphicx}
\usepackage{tabularx}
\usepackage{gensymb}
\usepackage{amsmath}
\usepackage{bm}
\DeclareUnicodeCharacter{2061}{}
\DeclareUnicodeCharacter{2212}{}
\DeclareUnicodeCharacter{0308}{}

\author{Lu Luo}
\author{Mahmoud RM Atalla}
\author{Simone Assali}
\author{Sebastian Koelling}
\author{Oussama Moutanabbir}
\affiliation[Polytechnique Montreal]
{Department of Engineering Physics, \'Ecole Polytechnique de Montr\'eal, Montr\'eal, C.P. 6079, Succ. Centre-Ville, Montr\'eal, Qu\'ebec, Canada H3C 3A7}

\title[An \textsf{achemso} demo]
  {Mid-infrared top-gated Ge/Ge$_{0.82}$Sn$_{0.18}$ nanowire phototransistors}
\begin{document}
\section{Abstract}
Achieving high crystalline quality Ge$_{1-x}$Sn$_{x}$ semiconductors at Sn content exceeding 10\% is quintessential to implementing the long sought-after silicon-compatible mid-infrared photonics. Herein, by using sub-20 nm Ge nanowires as compliant growth substrates, Ge$_{1-x}$Sn$_{x}$ alloys with a Sn content of 18\% exhibiting a high composition uniformity and crystallinity along a few micrometers in the nanowire growth direction were demonstrated. The measured bandgap energy of the obtained Ge/Ge$_{0.82}$Sn$_{0.18}$ core/shell nanowires is 0.322 eV enabling the mid-infrared photodetection with a cutoff wavelength of 3.9 $\mu$m. These narrow bandgap nanowires were also integrated into top-gated field-effect transistors and phototransistors. Depending on the gate design, these demonstrated transistors were found to exhibit either ambipolar or unipolar behavior with a subthreshold swing as low as 228 mV/decade measured at 85 K. Moreover,  varying the top gate voltage from  -1 V to 5 V yields nearly one order of magnitude increase in the photocurrent generated by the nanowire phototransistor under a 2330 nm illumination. This study shows that the core/shell nanowire architecture with a super thin core not only mitigates the challenges associated with strain buildup observed in thin films but also provides a promising platform for all-group IV mid-infrared photonics and nanoelectronics paving the way toward sensing and imaging applications.\\

\bigskip
\textbf{Keywords:} Mid-infrared photonics; Germanium-tin alloys; Silicon photonics; Nanowires; Sensing and imaging.
\newpage

\section{Introduction}
Group IV Ge$_{1-x}$Sn$_{x}$ alloys is an emerging class of narrow direct bandgap semiconductors that has recently been attracting a great deal of attention as a versatile platform for silicon-compatible mid-wave infrared (MWIR) photonics and optoelectronics.\cite{moutanabbir2021monolithic} The indirect-to-direct bandgap crossover can be achieved in these materials at a strain-dependent Sn content typically in the 6$\sim$10 at.\% range.\cite{gupta2013achieving,attiaoui2014,eckhardt2014indirect} However, controlling the growth of Ge$_{1-x}$Sn$_{x}$ above this critical content to cover the MWIR range has been facing major hurdles due to the low solid solubility of Sn in Ge (\textless 1 at.\%) and the large lattice mismatch between Ge and Sn ($\approx$14.7\%). These limitations can be mitigated, to a certain extent, through nonequilibrium epitaxial growth processes introducing strain-relaxed Ge$_{1-x}$Sn$_{x}$/Ge buffer layers.\cite{assali2018atomically,dou2018investigation} While lattice mismatch-induced high compressive strain can be partially relaxed by forming multiple step-graded layers, the highly defective bottom Ge$_{1-x}$Sn$_{x}$ layers can still severely degrade the device performance\cite{buca2022room,chretien2022room,zhou2020electrically,atalla2022high,atalla2023dark,tran2019si} in addition to limiting the thermal stability of the material.\cite{Nicolascgd2020,mukherjee2021atomic} To circumvent the harmful effects of these extended defects, nanoscale Ge$_{1-x}$Sn$_{x}$ material systems have been proposed as an alternative platform.\cite{doherty2020progress,WinNT} Indeed, Ge$_{1-x}$Sn$_{x}$ nanowires (NWs) were shown to provide an effective path to relax the lattice mismatch-induced compressive strain through the free sidewall facets without forming a high density of defects in the material.\cite{albani2018critical,kavanagh2012faster,lewis2017anomalous} Besides, these nanoscale materials also provide additional degrees of freedom to engineer and control light-matter interaction and the related absorption and detection.\cite{cao2009engineering,cao2010resonant,svensson2013diameter,AttiaouiPRA2021,Attiaouiadma2023} 

Ge$_{1-x}$Sn$_{x}$ NWs reported in recent studies were synthesized using various methods including the vapor-liquid-solid (VLS),\cite{biswas2016non,doherty2019one,doherty2018influence,seifner2019epitaxial} in-plane solution–liquid–solid (IPSLS),\cite{azrak2018growth,azrak2019low} microwave-assisted solution-liquid-solid growth,\cite{barth2015microwave,seifner2017pushing} and top-down etching.\cite{eustache2021smooth,attiaoui2021extended} However, these methods typically yield NWs with either low Sn content (below 10 at.\%) or irregular shapes thus limiting their potential for MWIR device development. In this regard, Ge/Ge$_{1-x}$Sn$_{x}$ core/shell NW growth emerged as a reliable approach to obtain nanoscale Ge$_{1-x}$Sn$_{x}$ alloys at a higher and more uniform Sn composition.\cite{meng2016core,assali2017growth,luo2023midinfrared,kim2023short} In this process, the Ge core acts as a compliant growth template, which facilitates the relaxation of the compressive strain, leading to an increased Sn content and higher uniformity\cite{luo2023midinfrared,assali2019strain,assali2020kinetic} thus enabling the development of innovative nanoscale MWIR emitters and detectors.\cite{luo2022extended,luo2023midinfrared,singh2022ge,kim2023short} For example, Ge/Ge$_{1-x}$Sn$_{x}$ core/shell NWs with a sub-20 nm Ge core and ${x}$ in the 0.06-0.18 range were recently demonstrated and integrated in photoconductive detectors allowing a tunable cutoff wavelength covering from 2.0 $\mu$m to 3.9 $\mu$m.\cite{luo2023midinfrared} Although these NW photoconductors are straightforward to fabricate and can provide relatively high photoresponsivity and photoconductive gain due to the extended carrier lifetime, their high dark current still remains a major burden. Moreover, both the NW small size and the low energy of MWIR photons limit the NW photodetector absorption efficiency,  thus making it harder to accurately detect weak signals. These limitations are further exacerbated in single NW photodetectors. Hence, it is highly critical to integrate individual NWs into photodetector configurations optimized to yield a reduced dark current. Such requirements can be met in device architectures such as junctionless phototransistors.\cite{roudsari2014junction,colinge2011junctionless} \\
A junctionless phototransistor simply involves the introduction of a gate contact to control the conductivity of the NW channel between the drain and the source without forming a PN junction thus eliminating the need for post-growth doping and annealing that can be complicated for Ge$_{1-x}$Sn$_{x}$-based materials due to their metastable nature. The gate can deplete the NW from its majority carriers by providing a potential barrier, thus effectively reducing the dark current. The prevalent configuration for NW phototransistors is the back-gated phototransistor, yet it exhibits a limited gate control.\cite{galluccio2020field,appenzeller2008toward} To simultaneously solve all of these challenges, in this work a top-gated field-effect transistors (FETs) were fabricated by using narrow bandgap Ge/Ge$_{0.82}$Sn$_{0.18}$ core/shell NWs and the phototransistor performance in the MWIR range was demonstrated. The FET devices exhibit either ambipolar or unipolar transistor behavior depending on the design of the gate which can fully or partially and asymmetrically cover the transistor channel. Subthreshold swing values as low as 833 mV/decade and 228 mV/decade were achieved at 300 K and 85 K, respectively. The power-dependent and gate-modulated photocurrent at 2330 nm illumination demonstrate a stable performance of the NW phototransistors. These results highlight the potential of Ge$_{1-x}$Sn$_{x}$ nanoscale alloys as versatile building blocks for group IV nanoelectronics and MWIR nanophotonics. \\
\begin{figure}[th]
    \centering
    \includegraphics[width=16cm]{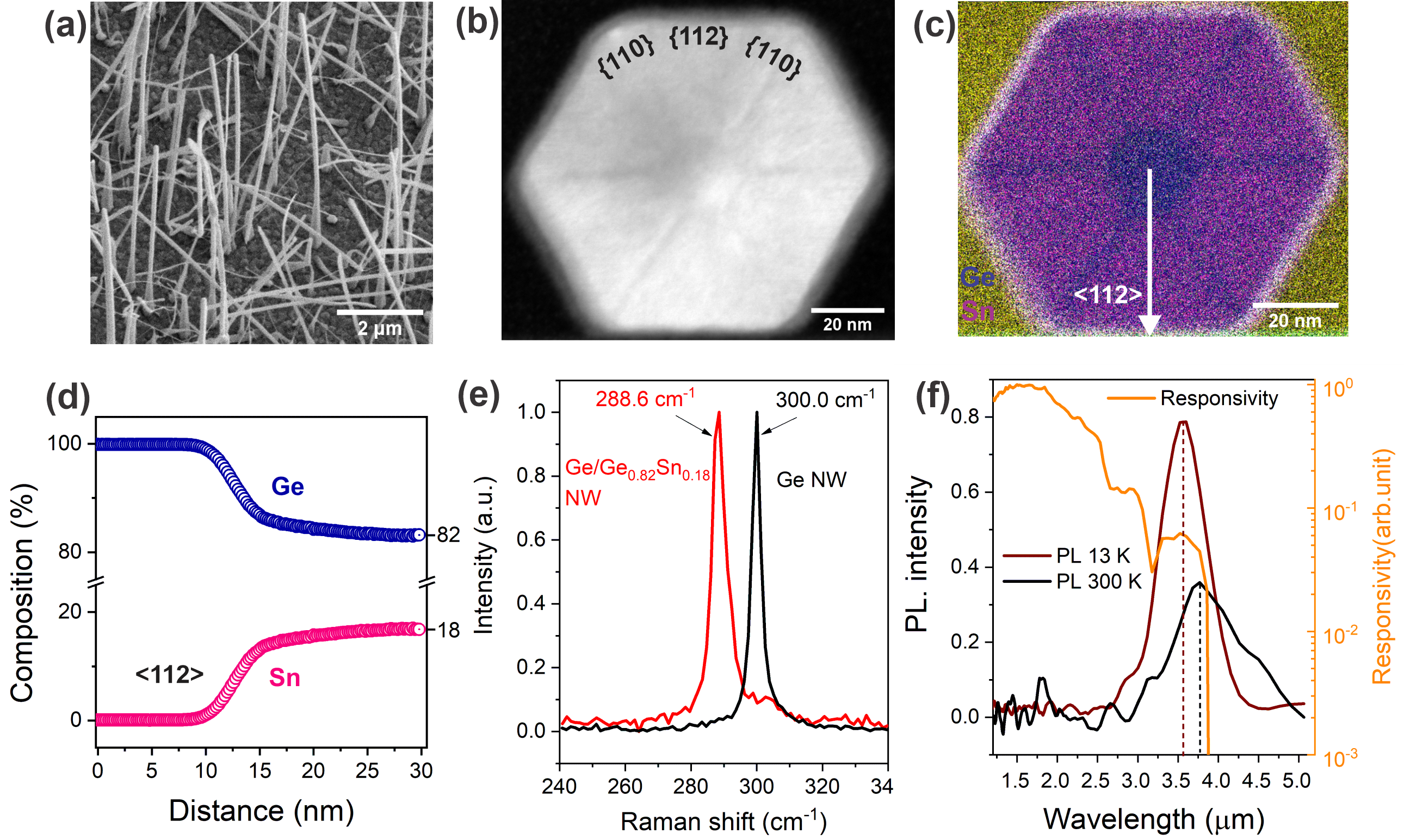}
    \caption{(a) SEM image of the as-grown Ge/Ge$_{0.82}$Sn$_{0.18}$ core/shell NWs. (b) Low magnification High-Angle Annular Dark-Field Scanning Transmission Electron Microscopy (HAADF-STEM) image showing the cross-section of the NW viewed along the $<111>$ direction. (c) Energy dispersive X-Ray (EDX) elemental mapping of the NW cross-section. (d) Ge and Sn composition profiles along the $<112>$ direction, indicated by the white arrow in (c), as measured by Atom Probe Tomography (APT). (e) Raman spectra of single Ge/Ge$_{0.82}$Sn$_{0.18}$ core/shell NW and Ge NW. (f) Photoluminescence spectra of the as-grown NW at 300 K and 85 K, and spectral responsivity of multiple NW device adapted from  Ref.\cite{luo2023midinfrared}}
    \label{fig.1}
\end{figure}
\section{Results}
Ge/Ge$_{0.82}$Sn$_{0.18}$ core/shell NWs were grown on Ge(111) substrates in a reduced-pressure chemical vapor deposition (CVD) reactor following two growth steps. First, Ge core NWs were grown at 340 $^{\circ}$C by the VLS process, where 20 nm-diameter Au particles and GeH$_4$ served as the catalyst and the precursor, respectively. Second, Ge$_{0.82}$Sn$_{0.18}$ shells with 35-40 nm in thickness were epitaxially grown at 260 $^{\circ}$C on the obtained Ge NWs with the supply of GeH$_4$ and SnCl$_4$ precursors. Fig. 1(a) shows a scanning electron microscopy (SEM) image of the as-grown NWs with diameters and lengths in the range of 90-100 nm and 5-7 $\mu$m, respectively. Fig. 1(b) presents the low magnification high-angle annular dark-field scanning transmission electron microscopy (HAADF-STEM) image showing the NW cross-section viewed along the $<111>$ direction. The energy dispersive X-ray (EDX) elemental map of the NW cross section is displayed in Fig. 1(c).  A clear core/shell structure is observed arising from the Z-contrast of the image in Fig. 1(b). In addition, a 6-fold Sn-depleted substructure along the corners of the hexagonal GeSn shell (especially in the EDX image) is also visible, which corresponds to the \{110\} crystal facets. This phenomenon is associated with facet-dependent capillarity diffusion due to the nonplanarity of the shell growth, which has been reported in group III-V-based NW heterostructures.\cite{zheng2013polarity,rudolph2013spontaneous} Ge and Sn composition profiles measured by atom probe tomography (APT) along the $<112>$ direction in the NW cross section are displayed in Fig. 1(d). The Sn composition increases rapidly from 0 to 15 at.\% in the first 5 nm of the shell before reaching the targeted Sn content of 18 at.\%. This rapid transition from Ge to Ge$_{0.82}$Sn$_{0.18}$, without a significant strain buildup in Ge$_{0.82}$Sn$_{0.18}$ as shown below, is peculiar to the core/shell growth exploiting the compliant nature of the sub-20 Ge NW.\cite{luo2023midinfrared}

Raman spectroscopy measurements were performed on single NWs using a 633 nm excitation laser having a 0.7 $\mu$m-diameter spot size. A representative set of the recorded Raman spectra is shown in Fig. 1(e). It is noticeable that Ge-Ge LO phonon mode is shifted from 300.0 cm$^{-1}$ in Ge NW to 288.6 cm$^{-1}$ in Ge/Ge$_{0.82}$Sn$_{0.18}$ core/shell NW suggesting that the latter are nearly fully relaxed. Indeed, the estimated residual strain is about -0.2\% in these core/shell NWs, which is significantly below the -1.3\% strain measured in Ge$_{1-x}$Sn$_{x}$ pseudomorphic layers at similar content.\cite{bouthillier2020decoupling,assali2018atomically}   

Photoluminescence (PL) measurements at 300 K and 13 K were recorded for as-grown Ge/Ge$_{0.82}$Sn$_{0.18}$ core/shell NWs, as shown in Fig. 1(f). The description of the PL setup can be found in Methods. This figure also displays the responsivity spectrum of a multiple NW photoconductive device. The behavior of the latter is discussed in detail in Ref.\cite{luo2023midinfrared}.  As expected,  the peak wavelength red-shifts from 3.56 $\mu$m at 13 K to 3.77 $\mu$m at 300 K. The reduced bandgap energy at the high temperature results from the increased electron–phonon interactions and lattice expansion while the broader PL spectra at room temperature can result from the enhanced electron–phonon coupling.\cite{zhang2017temperature,varshni1967temperature} Note that the PL peak wavelength at room temperature is close to the measured cutoff wavelength (3.85 $\mu$m) of the photoconductive devices, providing an estimated bandgap energy of 0.322 eV at 300 K. \\

\begin{figure}[th]
    \centering
    \includegraphics[width=16cm]{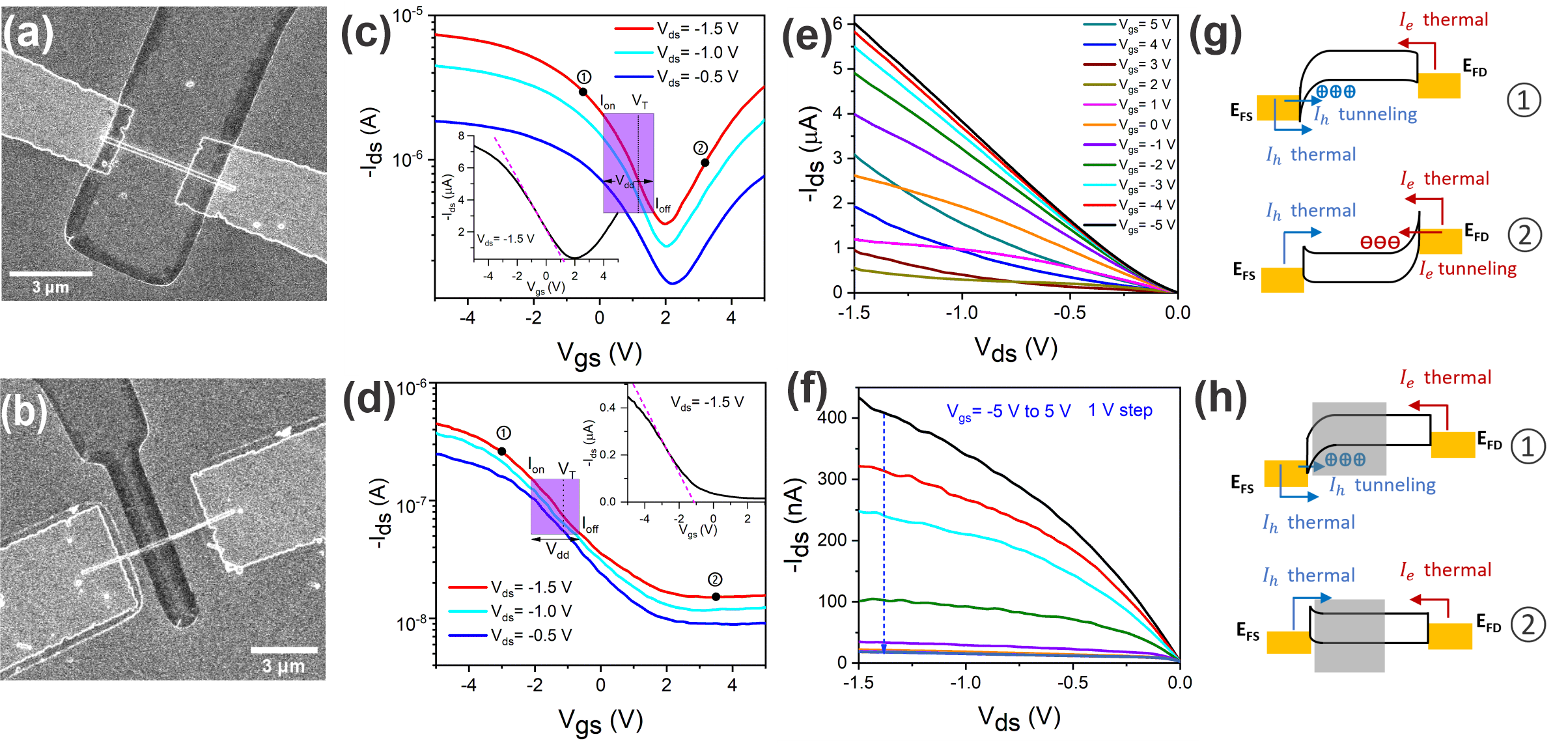}
    \caption{(a) and (b) Top view SEM images of devices A and B. (c) and (d) Logarithmic scale of I$_{ds}$-V$_{gs}$ curves for device A and device B, respectively, the insets are the linear scale of I$_{ds}$-V$_{gs}$ curve that was used to obtain the threshold voltages. (e) and (f) I$_{ds}$-V$_{ds}$ curves for device A and device B respectively. (g) and (h) Band bending diagram explaining the ambipolar and unipolar transistor behavior of device A and B, respectively.}
    \label{fig.2}
\end{figure}

Top-gated FETs were fabricated (see Methods) to investigate the optoelectronic performance of the Ge/Ge$_{0.82}$Sn$_{0.18}$ core/shell NWs. 20 nm-thick HfO$_2$ was deposited by atomic layer deposition (ALD) as the gate oxide. Two types of devices were fabricated: one with a gate contact fully covering the NW (device A), and the other with a gate contact partially and asymmetrically covering the NW (device B). Fig. 2(a) and 2(b) are representative SEM images of device A and device B, where the gate length L is 2.7 $\mu$m and 1.3 $\mu$m, respectively. The typical I$_{ds}$-V$_{gs}$ curves of the two NW FETs at V$_{ds}$ = -0.5 V, -1.0 V, and -1.5 V are shown in Fig. 2(c) and 2(d). It is worth noting that the drain current I$_{ds}$ in device A increases with both negative and positive gate voltages indicating p-type and n-type transfer properties with the p-type transfer as dominant. However, this ambipolar transistor behavior can be suppressed by the asymmetrical gate design, as the drain current I$_{ds}$ in device B only increases with negative gate voltage suggesting a p-type FET behavior. The I$_{ds}$-V$_{ds}$ output characteristic curves of the two devices with V$_{gs}$ changing from -5 V to 5 V are exhibited in Fig. 2(e) and 2(f), showing the ambipolar and unipolar behavior. The observed ambipolarity arises from the narrow bandgap of Ge$_{0.82}$Sn$_{0.18}$ NW, which is also a common behavior in InSb NWs\cite{dalelkhan2020ambipolar} and black phosphorus FETs.\cite{du2014device} 

In a narrow bandgap material, the band bending can create a small triangular potential barrier in the metal-semiconductor interface where the electrons can easily tunnel through, therefore yielding a tunneling current. Besides, the electrons can be thermally excited through the Schottky barrier to form a thermal emission current.\cite{hutin2009schottky,penumatcha2015analysing} Therefore, the total drain current I$_{ds}$ is the sum of the hole current I$_h$ and electron current I$_e$ including the contributions of both tunneling and thermionic emission for each carrier type. The thermionic emission current will always exist in a Schottky barrier, while the tunneling current becomes noticeable only when the triangular barrier width is sufficiently narrow. Fig. 2(g) exhibits the band diagrams at point 1 and point 2, marked in Fig. 2(c), describing the origin of the ambipolar transistor behavior of device A. Fig. 2(h) shows the band diagrams at point 1 and point 2, indicated in Fig. 2(d), illustrating the unipolar transistor behavior of device B. For the top-gated transistor with the channel fully covered by the gate contact, tunneling dominates the current when the gate voltage is raised positively (n-channel) or negatively (p-channel) due to the bent band at the metal-drain and metal-source interfaces. The unipolar transfer behavior in device B is the result of the asymmetrical gate control. Due to the absence of the gate control, Fig. 2(h) shows that the band of the larger ungated region (right side) is more difficult to bend than that of the shorter ungated region (left side). Thus, the thermal emission current dominates on the drain side where not covered by the gate contact. When the gate voltage increases toward positive values, the electron tunneling current is blocked by the relatively thick barrier at the drain-metal interface. 

The on/off ratio and subthreshold swing (SS) captures the off-state leakage and the switching speed, which are important parameters to evaluate the transistor performance. By following the method in \cite{chau2005benchmarking,xiang2006ge}, the on  (I$_{on}$) and off (I$_{off}$) currents for a FET should be the values measured at V$_g$=V$_T$-0.7V$_{dd}$ and V$_g$=V$_T$+0.3V$_{dd}$ in the I$_{ds}$-V$_{gs}$ curve, respectively. V$_{dd}$ is the power supply voltage which is equal to 1.5 V in this study, and V$_T$ is the threshold voltage which is extracted using the standard peak transconductance method.\cite{schroder2015semiconductor} V$_T$ for device A and device B is estimated at 1.2 V and -1.1 V and their on/off ratios are 4.6 and 2.8, respectively. 

The SS values, defined as $R = [\frac{d(log_{10}{I_d})⁡}{dVg}]^{-1}$, were found to be respectively 1860 mV/decade and 3300 mV/decade for device A and device B, which are similar to those reported for InSb NWs\cite{dalelkhan2020ambipolar,candebat2009insb} InAs NW,\cite{li2018improving} and black phosphorus FETs.\cite{du2014device} The low on/off ratio and the weak SS in both Ge/Ge$_{0.82}$Sn$_{0.18}$ NW FETs result from the high off-state current. The high off-state current in a narrow bandgap NW FET can be attributed to several reasons including the gate leakage current, thermionic emission current (as mentioned above), trap-assisted tunneling (TAT) current, and ambipolar current.\cite{zhu2013low,zhu2013reliability} However, the first reason can be excluded because the measured gate leakage is less than 17 pA (as outlined in the Supplementary Information, Section 1), which is close to the noise level. This means that a uniform and conformal ALD-deposited dielectric layer was formed on the NW FETs, as shown in Fig. S2(a) and S2(b). Note that the off-state current in device B is $\sim$3 to 10 times less than that of device A, indicating that the asymmetrical gate design can shrink the off-state current. Indeed, the high off-state current in device A is attributed to the ambipolar behavior. The same phenomenon was reported in InAs NW and NW-based tunnel field-effect transistors.\cite{yang2018suppressing,verhulst2007tunnel} Both the thermionic emission and the TAT current are sensitive to temperature. The TAT current results from defects at the drain/channel interface, which can create energy levels within the bandgap and assist electrons in tunneling through the barrier.\cite{zhu2013reliability} 
To evaluate the temperature dependence of the off-state current of Ge/Ge$_{0.82}$Sn$_{0.18}$ NW FETs, transport measurements at low temperature were conducted in a cryogenic probe station (see Methods for details). Fig. 3(a) and 3(b) displays the transfer characteristic curves for the two devices measured at 300 K and 85 K at a drain voltage of -0.2 V and -0.5 V for the gate voltage ranging from -5 V to 5 V and -2 V to 8 V, respectively. The off-state current in both devices was largely reduced as expected, leading to an increased on/off ratio and a reduced SS value. Note that the devices in Fig. 3 are not the same devices as in Fig. 2.

\begin{figure}[th]
    \centering
    \includegraphics[width=16cm]{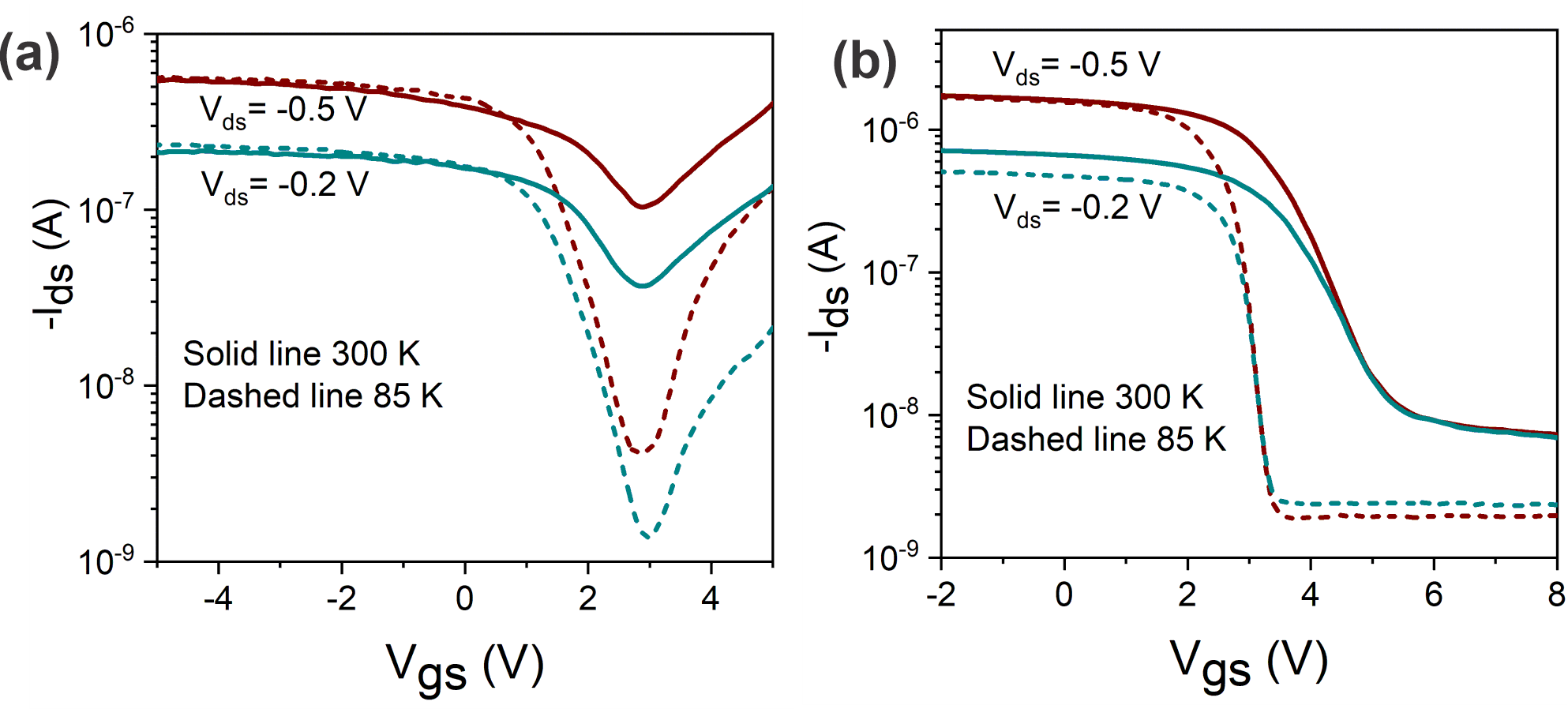}
    \caption{(a) I$_{ds}$-V$_{gs}$ curves for device A at 300 K and 85 K. (b) I$_{ds}$-V$_{gs}$ curves for device B at 300 K and 85 K}
    \label{fig.3}
\end{figure}

Following the calculation method mentioned above, the I$_{on}$/I$_{off}$ ratio of device A and device B were found to increase from 1.57 and 3.5 at 300 K to 3.87 and 51.1 at 85 K, respectively. Simultaneously, both devices show a decrease in the SS value as the temperature is cooled down to 85 K from 2247 mV/decade to 664 mV/decade in device A and from 833 mV/decade to 228 mV/decade in device B. The SS is expected to change linearly with temperature following $(ln10)({kT\over q})(\frac{C_{ox}+C_d}{C_d})$, where C$_d$ and C$_{ox}$ are depletion capacitance and gate oxide capacitance, and kT/q is the thermal voltage. The ratio of SS change in both devices is very close to the ratio of temperature change, which is 3.53. 

\begin{figure}[ht]
    \centering
    \includegraphics[width=14cm]{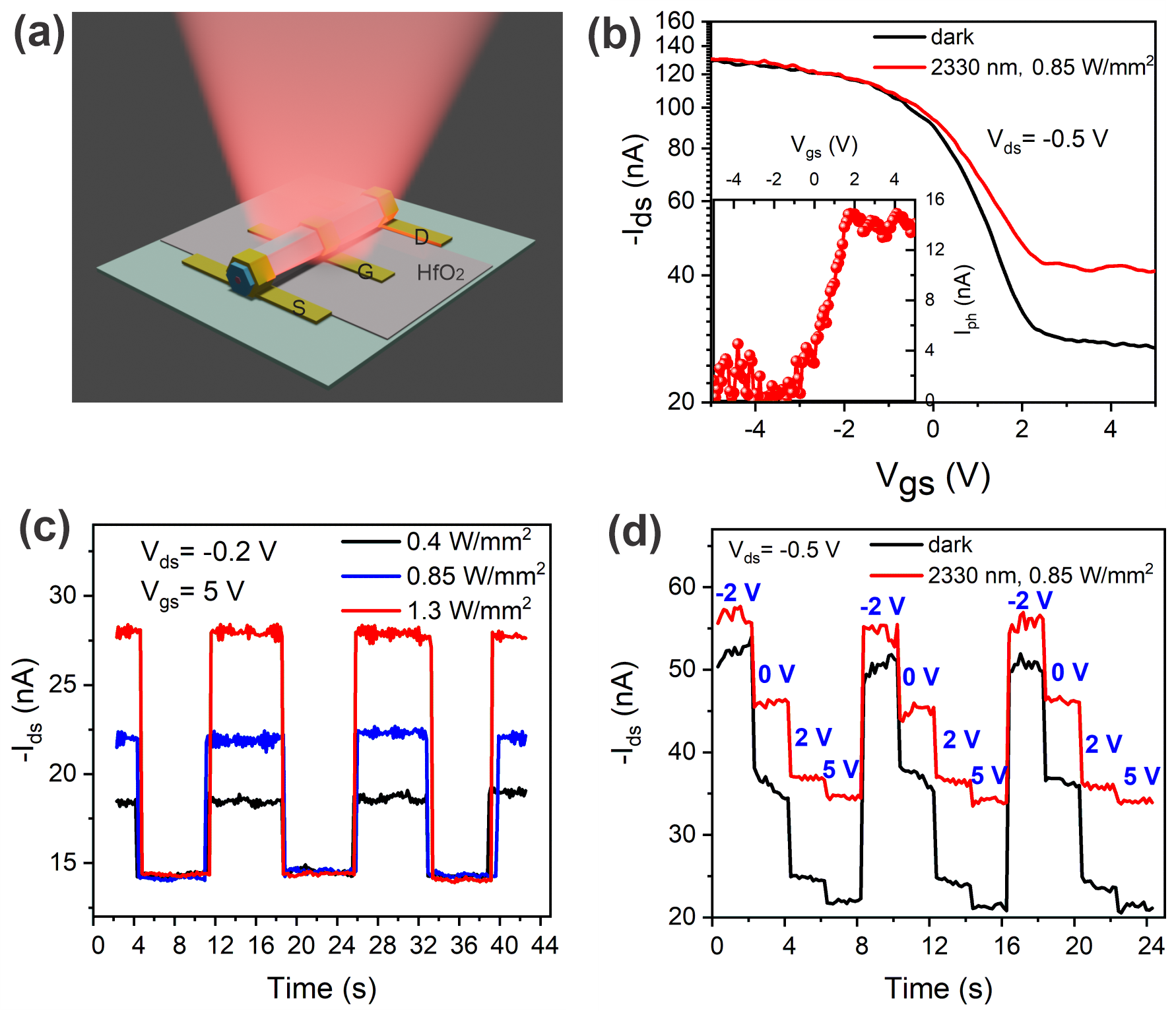}
    \caption{(a) Schematic illustration of the Ge/Ge$_{0.82}$Sn$_{0.18}$ core/shell NW top-gated phototransistor. (b) I$_{ds}$-V$_{gs}$ curves of the top-gated phototransistor under dark condition and 2330 nm illumination. The inset shows the measured photocurrent vs. the gate voltage. (c) Photocurrent of the top-gated phototransistor under 2330 nm laser at different power densities. (d) Time-resolved drain current at four different gate voltages under dark and 2330 nm illumination conditions.}
    \label{fig.4}
\end{figure}

The gate voltage modulated photoresponse performance of the Ge/Ge$_{0.82}$Sn$_{0.18}$ NW phototransistor was also investigated at room temperature. The schematic of the phototransistor is illustrated in Fig. 4(a), where a 2330 nm laser diode was used as the light source. Fig. 4(b) depicts the I$_{ds}$-V$_{gs}$ curves with and without illumination when the drain voltage V$_{ds}$ is -0.5 V, and the inset displays the photocurrent I$_{ph}$ versus V$_{gs}$ extracted from Fig. 4(b). A small photocurrent oscillating around 2 nA is detected when the gate voltage is below -1 V then increases gradually until it reaches saturation at 14 nA with the gate voltage increased from -1 V to 5 V. Fig. 4(c) shows the transient photocurrent at varied light intensities under the conditions of V$_{ds}$= -0.2 V and V$_{gs}$= 5 V. In Fig. 4(d), the transient drain current with and without light at different gate voltages are shown for V$_{ds}$= -0.5 V. The gate-modulated photocurrent in this figure is in good agreement with the behavior in Fig. 4(b). The power-dependent and gated-modulated photoresponse in Fig. 4(c) and 4(d) indicate a stable performance of the NW top-gated mid-infrared phototransistors. In fact, the top-gated NW FETs show a photoresponse and a gate modulation performance comparable to those of n-type InAs back-gated FET.\cite{zhang2020surface} Additionally, in comparison to the side-gated NW FETs,\cite{yang2020ferroelectric} the top-gated NW FETs demonstrated in this work exhibit at least a 50 \% increase in the on/off ratio and a 7-fold enhancement in photocurrent while being under 10 times lower light power density. \\

In conclusion, nanoscale Ge$_{0.82}$Sn$_{0.18}$ alloys with high crystallinity and content uniformity were achieved using sub-20 nm Ge NWs as compliant growth substrates. The obtained Ge/Ge$_{0.82}$Sn$_{0.18}$ core/shell NWs have a bandgap energy of 0.322 eV allowing the mid-infrared photodetection with a cutoff wavelength of 3.9 $\mu$m. The grown NWs were also introduced in top-gated FETs exhibiting a strong ambipolar behavior with a small I$_{on}$/I$_{off}$ ratio (less than 10). The ambipolar behavior and the off-state current of Ge/Ge$_{0.82}$Sn$_{0.18}$ NW FETs can be reduced by introducing an asymmetrical top gate contact. Power-dependent and gate-modulated photoresponse under a 2330 nm laser illumination indicated a stable gate control in the NW top-gated phototransistors, which yielded at least a 50 \% increase in the on/off ratio and a 7-fold enhancement in photocurrent under 10 times lower light power density as compared to the side-gated NW FETs. These results hint at the potential of nanoscale Ge$_{1-x}$Sn$_{x}$ semiconductors for all-group IV mid-infrared optoelectronics.

\section{Methods}
\textbf{Fabrication of top-gated NW device.} The as-grown NWs were drop-casted onto a Si substrate with a 100 nm-thick SiO$_2$ layer. Then methyl methacrylate (MMA) and poly methyl methacrylate (PMMA) were spin-coated on the SiO$_2$/Si substrate. EBL (Raith 150) was used to define drain and source patterns. Afterward, the sample was dipped into 5 \% HCl solution for 30 s to remove the native oxide. The sample was subsequently transferred to an E-beam evaporator chamber quickly to deposit Cr/Au (5/150 nm). After the first lift-off, 20 nm of HfO$_2$ layer was deposited on the sample by ALD followed by a second EBL, metal deposition, and the liftoff procedures to form the top-gate contacts.

\textbf{Photoluminescence setup.} The NW sample was fixed on the sample holder inside the cryogenic chamber. A helium closed loop was installed to cool down the chamber. The excitation was carried out using an 808 nm laser focused on the sample using a 25x objective. The emission coming from the sample was fed into a Fourier-transform infrared spectrometer. 

\textbf{Low-temperature measurement setup.} A cryogenic probe station from Janis Research Company was used to carry out low-temperature transport measurements. The chamber was evacuated by a mechanical pump, where the liquid Ni was introduced once the pressure reached 10$^{−8}$ Psi. The temperature in the chamber was controlled by a PID controller.

\begin{acknowledgement}
The authors thank J. Bouchard for the technical support with the CVD system.  O.M. acknowledges support from NSERC Canada (Discovery, SPG, and CRD Grants), Canada Research Chairs, Canada Foundation for Innovation, Mitacs, PRIMA Qu\'ebec, Defence Canada (Innovation for Defence Excellence and Security, IDEaS), the European Union’s Horizon Europe research and innovation programme under grant agreement No 101070700 (MIRAQLS), the US Army Research Office Grant No. W911NF-22-1-0277, and the Air Force Office of Scientific and Research Grant No. FA9550-23-1-0763. 

\end{acknowledgement}




\bibliography{achemso-demo}

\end{document}